\documentclass[english,aps,floats,twocolumn,showpacs,nofootinbib]{revtex4}
\usepackage{pslatex}
\usepackage[T1]{fontenc}
\usepackage[latin1]{inputenc}
\usepackage{graphicx}
\usepackage{epsfig}

\usepackage{calc}
\usepackage{ifthen}

{\newcommand{\lsim}{\mbox{\raisebox{-.6ex}{~$\stackrel{<}{\sim}$~}}}
{\newcommand{\gsim}{\mbox{\raisebox{-.6ex}{~$\stackrel{>}{\sim}$~}}}
{
\newcommand{\be}{\begin{equation}}
\newcommand{\ee}{\end{equation}}
\newcommand{\bea}{\begin{eqnarray}}
\newcommand{\eea}{\end{eqnarray}}

\begin{document}
\title{Absolute electron and positron fluxes from PAMELA/Fermi and Dark Matter}
\author{C. Bal\'azs$^{1}$}
\author{N. Sahu$^{2}$}
\author{A. Mazumdar$^{2,3}$}
\affiliation{
$^{1}$School of Physics, Monash University, Melbourne Victoria 3800, Australia\\
$^{2}$Physics Department, University of Lancaster, Lancaster LA1 4YB, UK\\
$^{3}$Niels Bohr Institute, Copenhagen University, Blegdamsvej-17, DK-2100, Denmark}

\begin{abstract}
We extract the positron and electron fluxes in the energy range 10 - 100 GeV by combining 
the recent data from PAMELA and Fermi LAT. The {\it absolute positron and electron} fluxes 
thus obtained are found to obey the power laws: $E^{-2.65}$ and $E^{-3.06}$ respectively, which 
can be confirmed by the upcoming data from PAMELA. The positron flux appears to indicate an excess 
at energies $E\gsim 50$ GeV even if the uncertainty in the secondary positron flux is added to 
the Galactic positron background. This leaves enough motivation for considering new physics, such 
as annihilation or decay of dark matter, as the origin of positron excess in the cosmic rays.

\end{abstract}
\maketitle

\section{Introduction}

Recent results from the Fermi Large Area Telescope (LAT)~\cite{fermi} indicate an excess of the electron 
plus positron flux ($\Phi_{e^-} + \Phi_{e^+}$) at energies above 100 GeV. This also confirms the
earlier results from ATIC~\cite{atic} and PPB-BETS~\cite{ppb_bets}, apart from the peak
at around 600 GeV in the ATIC data. The HESS~\cite{hess} collaboration also reported an excess of 
$\Phi_{e^-} +\Phi_{e^+}$ above 340 GeV, confirming their previous results~\cite{hess_past}. Meanwhile 
PAMELA data~\cite{pamela_positron} have shown an excess 
in ($\Phi_{e^+}/\Phi_{e^-}+\Phi_{e^+})$, compared to the Galactic background at 10-100 GeV. PAMELA also confirms the
earlier results from HEAT~\cite{heat_data} and AMS~\cite{ams_data}. Somewhat surprisingly, PAMELA did
not find any antiproton excess below 100 GeV ~\cite{pamela_proton}.

Although standard astrophysical sources may be able to account for the 
anomaly~\cite{pulsar}, the positron excess at PAMELA and the electron plus positron 
flux of Fermi have caused a lot of excitement being interpreted as indirect detection 
of dark matter (DM)~\cite{neutralino_clump,FermiDM}. If dark mater couples to standard model (SM) 
particles then the annihilation or decay of DM could, indeed, be the origin of cosmic ray anomalies 
observed by PAMELA and Fermi. The annihilation or decay of DM produces an equal number of particles 
(electrons and/or protons) and antiparticles (positrons and/or antiprotons), which 
could form a significant component of the observed cosmic rays. 
Since the background matter fluxes in the Galactic medium are at least one order of 
magnitude larger than the antiparticle fluxes, the DM signal is better observable in 
the Galactic antimatter fluxes.

Currently the antiproton flux up to 100 GeV in the PAMELA data is consistent with the 
Galactic background. However, there is an approximately 10\% excess in the PAMELA 
$\left( \Phi_{e^+}/(\Phi_{e^-}+\Phi_{e^+}) \right)$ data over the background. 
Since this excess is not given in terms of the absolute positron flux, it 
gives rise to various ambiguities~\cite{delhayaetal1}. 
The main source of these ambiguities is that the excess of the positron flux, 
as given by the PAMELA collaboration, crucially depends on the uncertainty
in the background of the electron as well as the positron 
fluxes~\cite{moskalenko&strong:astro1998, delhayaetal2, delhayaetal3}. 
Therefore it is not apparent that PAMELA implies a statistically significant 
positron excess until one shows it in terms of an absolute positron flux.

In this paper we make an attempt to disentangle the absolute positron flux up to an
energy of 100 GeV by combining the current data from PAMELA and Fermi. We quantify
the excess of the absolute positron flux after discussing the uncertainty in the 
secondary positron background flux~\cite{delhayaetal2,delhayaetal3}. It is shown that
the combined PAMELA and Fermi data seem to indicate an excess in the absolute positron
flux for $E\gsim 50$ GeV. We then demonstrate the compatibility of this excess with the 
annihilating DM scenario, and we compare this scenario with the current Fermi data. Moreover, 
we find that the absolute positron and electron fluxes admit power law spectra 
of $E^{-2.65}$ and $E^{-3.06}$ respectively.

\section{Absolute Positron Flux and Background}
So far no experiment has provided the absolute magnitude of the Galactic 
positron flux. Below 100 GeV the excess in the PAMELA data is about 10\%. 
On the other hand, the electron plus positron flux at Fermi does not seem to 
indicate any excess below 100 GeV. In order to accept the positron excess 
interpretation of PAMELA, it is crucial to show an excess in the absolute 
magnitude of the positron flux itself.

To decisively settle this issue, we combine the data from PAMELA and Fermi 
and extract the absolute positron and electron fluxes as:
\bea
\Phi_{e^+}  &=& \left( \frac{\Phi_{e^+}}{\Phi_{e^-}+\Phi_{e^+}} \right)_{\rm PAMELA} \times
\left( \Phi_{e^-}+\Phi_{e^+} \right)_{\rm Fermi} , \nonumber\\
\Phi_{e^-} &=& \left( \Phi_{e^-}+\Phi_{e^+} \right)_{\rm Fermi} - \Phi_{e^+} .
\label{absolute_fluxes}
\eea
In order to utilize Eq. (\ref{absolute_fluxes}), we need $\Phi_{e^-+e^+}$ from Fermi  
and $\Phi_{e^+}/(\Phi_{e^-} +\Phi_{e^+})$ by PAMELA in the same energy bin, so that the combined 
central values and uncertainties can be evaluated at a given energy $E$. 
To this end for each energy bin of the Fermi data we interpolate the PAMELA data.
We evaluate the uncertainty of the absolute positron and electron fluxes as: 
\bea
(\delta \Phi_{e^+})^2 = && \!\!\!\!\!\!\!\! (\Phi_{e^+})^2 \left(
  \left( \frac{ \delta(\Phi_{e^+}/(\Phi_{e^-}+\Phi_{e^+})) }{ \Phi_{e^+}/(\Phi_{e^-}+\Phi_{e^+}) } \right)^2 + \right.
  \nonumber \\ && \left. ~~~~~~~~~~~~
  \left( \frac{ \delta(\Phi_{e^-}+\Phi_{e^+}) }{ \Phi_{e^-}+\Phi_{e^+} } \right)^2 \right) , 
  \nonumber \\
(\delta \Phi_{e^-})^2 = && \!\!\!\!\!\!\!\! \left( \delta \Phi_{e^-+e^+} \right)^2 + 
\left( \delta \Phi_{e^+} \right)^2 .
\eea

\begin{figure}
\begin{center}
\epsfig{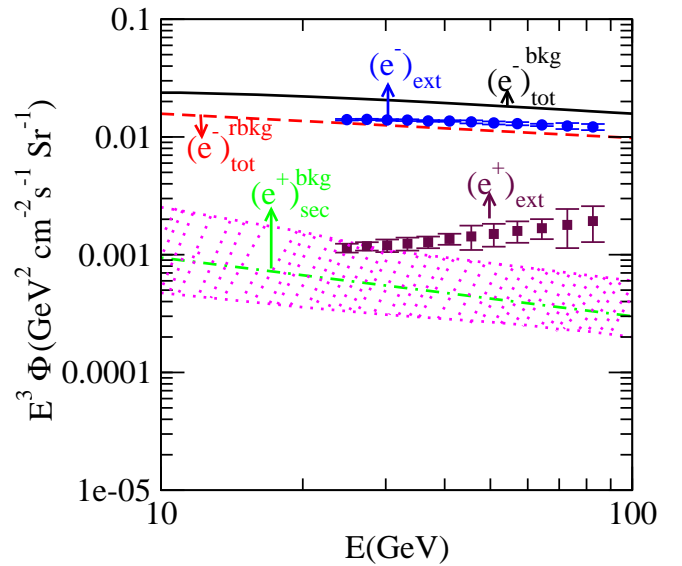}
\caption{The total (primary plus secondary) background electron flux (black-solid line),
the total reduced background electron flux (red-dashed line) and the positron background
(green dot-dashed line) with propagation uncertainty (pink band) are shown in the energy
range 10-100 GeV. Electron (blue-circle) and positron (maroon-square) fluxes, 
extracted from PAMELA and Fermi, are also shown with respect to their backgrounds. The 
power law fitting of positron and electron fluxes are found to be $E^{-2.56}$ and $E^{-3.06}$ 
respectively.}
\label{figure-1}
\end{center}
\end{figure}

Whether there is an excess or not in these combined fluxes will depend on the  
background expectations. The Galactic positron background fluxes were recently examined in 
Ref.s~\cite{moskalenko&strong:astro1998,delhayaetal2,delhayaetal3}. 
The majority of positrons in our galaxy are produced from scatterings of cosmic-ray 
protons with the interstellar medium. The positrons thus produced from proton-proton collisions 
provide background for the positrons produced from the annihilation or decay of DM. Therefore the background 
positrons in the Galactic medium are always secondary and can be parameterized as~\cite{baltz&edsjo:prd1998}:
\be
\Phi_{\rm sec,\; e^+}^{\rm bkg} = \frac{4.5 \epsilon^{0.7}}{1+650 \epsilon^{2.3}+1500 \epsilon^{4.2}}
{\rm GeV^{-1} cm^{-2} s^{-1} sr^{-1} }\,,
\label{background_fluxes}
\ee
where the dimensionless parameter is $\epsilon=E/$(1 GeV). However, there is a large uncertainty in the
secondary positron fluxes coming from cosmic ray propagation~\cite{delhayaetal1}, as shown in
FIG. \ref{figure-1}. In the Galactic medium the flux of the primary and secondary electrons is about
an order of magnitude larger than that of the positrons.  These electron fluxes are parameterized 
as~\cite{baltz&edsjo:prd1998}:
\bea
\Phi_{\rm prim,\; e^-}^{\rm bkg} = && \!\!\!\!\!\!\!\! \frac{0.16 \epsilon^{-1.1}}{1+11 \epsilon^{0.9}+3.2\epsilon^{2.15}}
{\rm GeV^{-1} cm^{-2} s^{-1} sr^{-1}}\nonumber\\
\Phi_{\rm sec,\; e^-}^{\rm bkg} = && \!\!\!\!\!\!\!\! \frac{0.70\epsilon^{0.7}}{1+11\epsilon^{1.5}+600 \epsilon^{2.9}
+580\epsilon^{4.2}} {\rm GeV^{-1} cm^{-2}s^{-1} sr^{-1}}\,. \nonumber \\
\eea
In FIG.~\ref{figure-1} we have shown the absolute positron and electron fluxes: $\Phi_{e^+}$ and $\Phi_{e^-}$, 
extracted from PAMELA and Fermi, up to 100 GeV. The corresponding error bars are also shown.
From FIG.~\ref{figure-1} it can be seen that the extracted positron flux
exhibits a clear excess with respect to the background for $E\gsim 50$~GeV. 
There is a minor excess of electron plus positron fluxes if we assume a 
10\% reduced background:
\be
\Phi_{e^-+e^+}^{\rm rbkg}=0.0253 \epsilon^{-3.206} {\rm GeV^{-1} cm^{-2} s^{-1} sr^{-1}}\,.
\label{reduced_ebg}
\ee
The reduced background can be thought of as uncertainty in the current estimation of 
electron flux~\cite{moskalenko&strong:astro1998}. Here we compare this reduced background 
with the $e^-+e^+$ spectrum of Fermi.

\section{Positron Excesses from DM Annihilation}
Annihilation of DM produces equally positrons and electrons in the Galactic medium
which can be a significant component of cosmic rays. However, as we discussed, the 
background electron flux in the Galactic medium is much larger than the background positron flux. 
Therefore finding a DM signal, if any, in the Galactic positron flux is easier than finding one 
in the electron flux. Thus, in what follows, we will focus on the positron flux in the cosmic rays.

Recent studies have shown that the $e^- + e^+$ spectrum of Fermi and the excess at PAMELA, without 
an excess in antiprotons, can simultaneously be explained with TeV scale DM, which annihilates to
$\mu^+ \mu^-$ and $\tau^+\tau^-$~\cite{strumiaetal,bergosrometal}. This explanation requires either 
the local density of DM or its annihilation cross section to be considerably higher than typically expected.
This enhancement is referred to as the ``boost factor". The origin of such a boost factor
could be astrophysical: large inhomogeneities, caused by merging sub-structures for example, could 
enhance the local density. Alternatively, assuming a homogeneous distribution of dark matter in the
galaxy, its annihilation cross section to $\mu^+ \mu^-$ and $\tau^+ \tau^-$ pairs at present could 
be significantly larger than that of the typical thermal relic, which is a few times $10^{-26}$
${\rm cm}^3/{\rm s} $~\cite{cireli:08}. The boost might arise from a combination of astro- and particle 
physics effects. 
Several
possible explanations for the boost factor have been proposed, in particular Sommerfeld
enhancement of the dark matter annihilation cross-section~\cite{cireli:08,winer:2008,barger:2008,
arkani_hamed:08,dutta:08,Bergstrom:2009ib}, non-thermal dark matter from thermal relic 
decay~\cite{fairbairn:2008,
zurek:08,kim:2008,spitzer:2008,Zant:2009sv,bhaskar:09,mcdonald2009,sahu2009,zhang_2009}, annihilation of 
thermal relic by resonance~\cite{ibeetal,pran_nath_2008} and so on. If the boost factor originates from
particle physics then its value is model dependent, but it should be constrained by the
Big-Bang Nucleosynthesis (BBN)~\cite{kohri:09}, gamma ray and radio observations~\cite{radiotest}, 
diffuse gamma ray background~\cite{diffuse_gammaray} and more severe constraints from gamma rays produced 
by inverse Compton scattering of the energetic electrons and positrons from DM annihilations~\cite{cirelli_2009}.

\subsection{Positron Propagation}
If the positrons are produced via the annihilation or decay of DM, then they travel in
the galaxy under the influence of a magnetic field which is assumed to be order of a few
micro-gauss. As a result the motion of positrons can be thought of as a random walk. The positron
flux in the vicinity of the solar system can be obtained by solving the diffusion
equation~\cite{delhayaetal2,hisanoetal:PRD2006,cireli&strumia:NPB2008}
\bea
\frac{\partial }{\partial t} f_{e^+}(E,\vec{r},t) &=& 
 K_{e^+}(E) \nabla^2 f_{e^+}(E,\vec{r},t) + \nonumber\\
& & \frac{\partial}{\partial t}[b(E) f_{e^+}(E,\vec{r},t)] + Q(E,\vec{r}) ,
\label{diffusion}
\eea
where $f_{e^+}(E,\vec{r})$ is the number density of positrons per unit energy,
$E$ is the energy of the positron, $K_{e^+}(E)$ is the diffusion constant, $b(E)$ is the
energy-loss rate and $Q(E,\vec{r})$ is the positron source term. The positron
source term $Q(E,\vec{r})$ due to DM annihilation is given by:
\be
Q(E,\vec{r})=\frac{1}{2} n_{\rm DM}^2(\vec{r})f_{\rm inj}^{{e^+}} ,
\label{source}
\ee
where the factor $1/2$ accounts the Majorana nature of DM, and the injection
spectrum can be given as
\be
f_{\rm inj}^{{e^+}} = \langle \sigma_{\rm DM} |v_{\rm rel}| \rangle
\frac{d N_{e^+}}{d E}\,.
\label{injection}
\ee
In the above equation the fragmentation function $d N_{e^+}/d E$ represents the
number of positrons with energy $E$ which are produced from the annihilation of DM.
We assume that the positrons are in steady state, i.e. $\partial f_{e^+}/\partial t=0$. Then from
Eq. (\ref{diffusion}), the positron flux in the vicinity of the solar system can be obtained in a
semi-analytical form~\cite{delhayaetal2,hisanoetal:PRD2006,cireli&strumia:NPB2008}
\bea
\Phi_{e^+} (E,\vec{r}_{\odot}) & = & \frac{v_{e^+}}{4\pi b(E)}(n_{\rm DM})_{\odot}^2
\langle \sigma_{\rm DM}|v_{\rm rel}| \rangle \times \nonumber\\
&& \int_E^{M_{\rm DM}} dE' \frac{dN_{e^+}}{dE'}I (\lambda_D(E,E'))\,,
\label{positron_flux}
\eea
where $\lambda_D(E,E')$ is the diffusion length from energy $E'$ to energy $E$ and $I(\lambda_D(E,E')$
is the halo function which is independent of particle physics. An analogous solution for the electron
flux can also be obtained.

The net positron flux in the galactic medium then can be given by
\be
(\Phi_{e^+})_{\rm Gal}=(\Phi_{e^+})_{\rm bkg} + \Phi_{e^+}(E,\vec{r}_{\odot}) .
\ee
The first term in the above equation is given by Eq. (\ref{background_fluxes}) while the
second term is given by Eq. (\ref{positron_flux}), which depends on various factors: $b(E)$,
$\lambda_D(E,E')$, $I (\lambda_D(E,E'))$, $v_{e^+}$, $(n_{\rm DM})_{\odot}$
and the injection spectrum $f_{\rm inj}^{{e^+}}$. The energy loss due to inverse Compton scattering
and synchrotron radiation with galactic magnetic field, described by $b(E)$, is determined by the photon
density and the magnetic field strength. Its value is taken to be $b(E)=10^{-16} \epsilon^2
{\rm GeV s}^{-1}$~\cite{baltz&edsjo:prd1998}. The number density of DM in the solar system is given by
\be
(n_{\rm DM})_{\odot}=\frac {\rho_\odot}{M_{\rm DM}} ,
\label{DM_density}
\ee
where $\rho_\odot\approx 0.3~{\rm GeV/cm^3}$. In the energy range we are interested in, the
value of $v_{e^+}$ is taken approximately to be $c$, the velocity of light. The values of
diffusion length $\lambda_D(E,E')$ and the corresponding halo function $I (\lambda_D(E,E'))$
are based on astrophysical assumptions~\cite{delhayaetal2,hisanoetal:PRD2006,cireli&strumia:NPB2008}.
By considering different heights of the galactic plane and different DM halo profiles the results may
vary slightly. In the following for the height of galactic plane we take $\lsim 4$ kpc, which is
referred to as "med" model~\cite{delhayaetal2,cireli&strumia:NPB2008}, and we have used the NFW DM
halo profile~\cite{NFW}
\be
\rho(r)=\rho_{\odot}\left( \frac{r_\odot}{r} \right)\left(\frac{1+ \left( \frac{r_\odot}{r_s} \right)
}{1+\left(\frac{r}{r_s} \right)} \right)^2 ,
\label{NFW-profile}
\ee
to determine the halo function $I (\lambda_D(E,E'))$, where $r_s\approx 20 {\rm kpc}$ and
$r_{\odot} \approx 8.5 {\rm kpc}$.
\begin{figure}[htbp]
\begin{center}
\epsfig{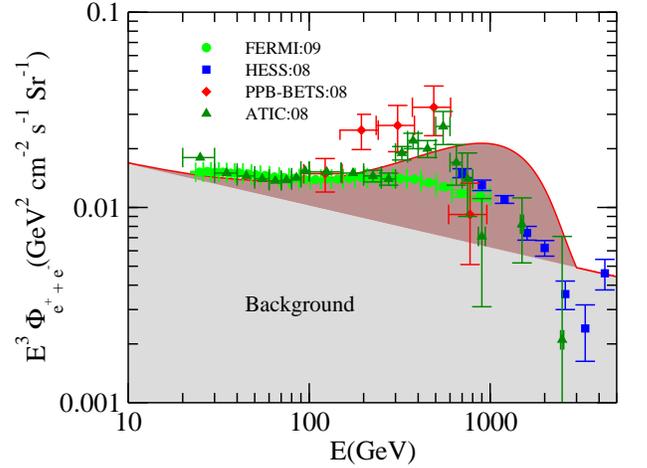}
\caption{$e^- + e^+$ flux from annihilation of DM to $\mu^+ \mu^-$. We have used
a boost factor of $5.3 \times 10^{3}$, a DM mass $M_{\rm DM}=3$ TeV, and the reduced background (\ref{reduced_ebg}).}
\label{DMtomuonpairs}
\end{center}
\end{figure}
\begin{figure}[htbp]
\begin{center}
\epsfig{file=fermi_pamela_mu.eps, width=0.45\textwidth}
\caption{Positron flux extracted from Fermi and PAMELA compared to the annihilation of DM to $\mu^+ \mu^-$. The required boost factor is $5.3 \times 10^{3}$ for a DM mass $M_{\rm DM}=3$ TeV.}
\label{DMtomuonpairs_pos}
\end{center}
\end{figure}
\begin{figure}[htbp]
\begin{center}
\epsfig{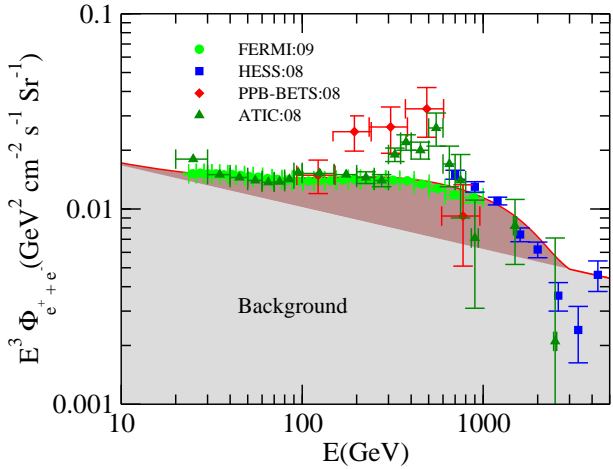}
\caption{$e^- + e^+$ flux from DM annihilation to $\tau^+ \tau^-$. The
boost factor is $6.7 \times 10^{3}$ and the DM mass ($M_{\rm DM}$) is 3 TeV. We have
used the reduced background (\ref{reduced_ebg}).}
\label{DMtotaupairs}
\end{center}
\end{figure}
\begin{figure}[htbp]
\begin{center}
\epsfig{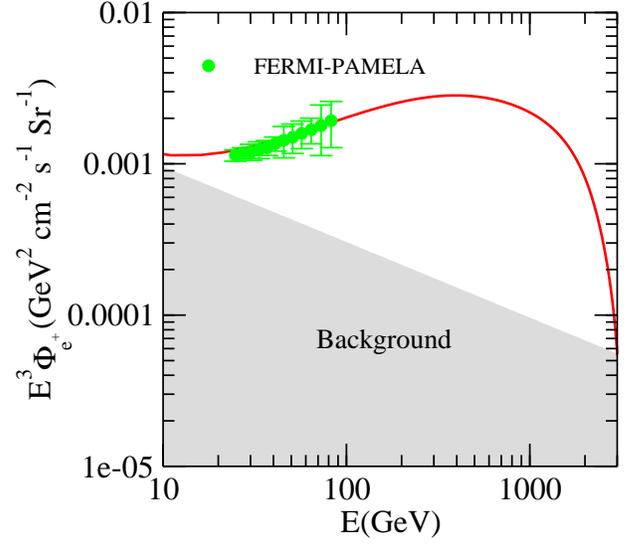}
\caption{Positron flux extracted from Fermi and PAMELA compared to the annihilation of DM to $\tau^+ \tau^-$.
The required boost factor is $5.3 \times 10^{3}$ for a DM mass $M_{\rm DM}=3$ TeV.}
\label{DMtotaupairs_pos}
\end{center}
\end{figure}
In FIG.s (\ref{DMtomuonpairs}) and (\ref{DMtotaupairs}) we have shown the the total
electron and positron fluxes from the annihilation of DM to muon and tau pairs respectively, 
plotted using DARKSUSY~\cite{darksusy}. Meanwhile, in FIG.s (\ref{DMtomuonpairs_pos}) 
and (\ref{DMtotaupairs_pos}) we have shown the extracted positron flux.  In the following 
we discuss a specific particle physics model where the DM annihilation to SM leptons can be 
enhanced through the Sommerfeld correction.

\subsection{A model for Sommerfeld enhanced DM annihilation cross-section}
As an illustrative example, we consider the model for Sommerfeld enhanced annihilation 
to muons which was proposed in Ref.~\cite{sahu2009,sahu2009_new}. The SM is extended by adding 
a hidden sector composed of three scalars $S$(1,0,3/2), $\chi(1,0,1)$ and $\phi(1,0,1)$, where 
the numbers inside the parenthesis are quantum numbers under gauge group $SU(2)_L\times U(1)_Y \times 
U(1)_{\rm hidden}$. The SM fields are neutral under $U(1)_{\rm hidden}$. The $U(1)_{\rm hidden}$ 
will be broken at around the electroweak scale to a surviving $Z_2$ symmetry under which $S$ is odd
while rest of the fields, including the SM fields, are even. As a result $S$ can be
a candidate for a DM~\cite{sdm1}. Several models have been considered in the literature~\cite{sdm}. 
Here we assume the mass of $S$ to be a few TeV, while the masses of
$\chi$ and $\phi$ to be of ${\cal O}(100)$GeV and ${\cal O}(100)$ MeV respectively.
The hidden sector is allowed to interact with the SM via the Higgs portal with
universal renormalisable couplings. The relevant Lagrangian is then given by
\bea
\mathcal{L} &\supseteq& f_{\rm portal} H^\dagger H \left(
S^\dagger S + \phi^\dagger \phi + \chi^\dagger \chi + \phi^\dagger \chi\right)\nonumber\\
&+& f_{S\phi} S^\dagger S \phi^\dagger \phi + f_{S\chi} S^\dagger S \chi^\dagger \chi
+ f_{S\chi\phi} S^\dagger S \chi^\dagger \phi + h.c.
\eea
Below 100 GeV $\chi$ acquires a vacuum expectation value (vev) and breaks $U(1)_{\rm hidden}$ to 
a surviving $Z_2$ symmetry under which $S$ is odd. It also gives a mixing between $H$ and $\phi$ through
the interaction term $H^\dagger H \phi^\dagger \chi$. As a result $\phi$ can potentially
annihilate to muon pairs. Since $\phi$ gets a mass through the vev of $H$,
the ${\cal O}(100)$ MeV scale mass of $\phi$ demands the universal
coupling of Higgs to hidden sector to be of ${\cal O} (10^{-6})$. However, we have
to make sure that with this coupling $S$ should be in thermal equilibrium above its
mass scale. This implies that
\be
f_{\rm portal} \gsim 8.36 \times 10^{-7} \left( \frac{M_S}{1 {\rm TeV}} \right)^{1/2}\,.
\ee
First $S$ will freeze-out at a temperature $T_S\sim M_S/20$. However, the corresponding
annihilation cross-section is known to be ${\cal O}(10^{-26})$ ${\rm cm}^3/{\rm sec}$. The
current annihilation of $S$ can be enhanced through the interaction: $S^\dagger S \chi^\dagger
\phi$. This interaction can generate an attractive force between S particles through
the exchange of $\phi$. The enhanced Sommerfeld annihilation cross-section then
requires $M_\phi \lsim \alpha M_S$~\cite{russell:08}, where $\alpha=\lambda^2/4\pi$, the effective
coupling $\lambda f_{S\chi\phi} \langle \chi \rangle /M_S$. This gives the constraint
on the coupling constant to be:
\be
f_{S\chi\phi}\gsim 0.5 \left(\frac{M_\phi}{200 {\rm MeV}} \right)^{1/2} \left(\frac{M_S}{1 {\rm TeV}} 
\right)^{1/2}
\left( \frac{100 GeV}{\langle \chi \rangle } \right)\,.
\ee
Thus we see that for $ f_{S\chi\phi}\gsim 0.5 $ we can get an enhanced annihilation cross-section to
explain the current anomalies at PAMELA and Fermi through $S$ annihilation to muons. The coupling
$f_{\rm portal} \ll f_{S\chi\phi}$ ensures that antiproton fluxes from $S^\dagger S$ annihilation
are suppressed.

\section{Conclusions}
In this paper we disentangled the absolute electron and positron fluxes by combining the current
data from PAMELA and Fermi. The electron and positron spectra are found to follow the power laws: 
$E^{-3.06}$ and $E^{-2.65}$ respectively. We showed that there is a clean excess of positron flux 
above 50 GeV even if the propagation uncertainty of positron is added to the background. 
This implies that we still have enough motivation for considering DM annihilation for the explanation of
current cosmic ray anomalies at PAMELA and Fermi. We then considered a variant of the model 
of Ref.~\cite{sahu2009} based on universal Higgs coupling to the hidden sector 
which can give rise to muon pairs from the annihilation of the dark matter particles.

\section{Acknowledgement } 
AM and NS were supported by the European Union through the Marie
Curie Research and Training Network ``UniverseNet" (MRTN-CT-2006-035863).  CB was 
funded by the Australian Research Council under Project ID DP0877916. NS would like to 
thank John McDonald and Kazunori Kohri for useful discussions.

\end{document}